\documentclass{article}
\usepackage{ijcai24}

\usepackage{times}
\usepackage{soul}
\usepackage{url}
\usepackage[hidelinks]{hyperref}
\usepackage[utf8]{inputenc}
\usepackage[small]{caption}
\usepackage{graphicx}
\usepackage{amsmath}
\usepackage{amsthm}
\usepackage{booktabs}
\usepackage{algorithm}
\usepackage{algorithmic}
\usepackage[switch]{lineno}
\usepackage{float}
\title{Unmasking Societal Biases in Respiratory Support for ICU Patients through Social Determinants of Health}

\author{
Mira Moukheiber$^1$
\and
Lama Moukheiber $^1$\and
Dana Moukheiber $^1$\And
Hyung-Chul Lee$^{2,}$\footnote{Corresponding author}\\
\affiliations
$^1$Massachusetts Institute of Technology\\
$^2$Seoul National University College of Medicine, Seoul National University Hospital, Department of Anesthesiology and Pain Medicine\\
\emails
\
vital@snu.ac.kr
}
\begin{document}

\maketitle

\begin{abstract}

In critical care settings, where precise and timely interventions are crucial for health outcomes, evaluating disparities in patient outcomes is important. Current approaches often fall short in comprehensively understanding and evaluating the impact of respiratory support interventions on individuals affected by social determinants of health. Attributes such as gender, race, and age are commonly assessed and essential, but provide only a partial view of the complexities faced by diverse populations. In this study, we focus on two clinically motivated tasks: prolonged mechanical ventilation and successful weaning. We also perform fairness audits on the models' predictions across demographic groups and social determinants of health to better understand the health inequities in respiratory interventions in the intensive care unit. We also release a temporal benchmark dataset, verified by clinical experts, to enable benchmarking of clinical respiratory intervention tasks.

\end{abstract}

\section{Introduction}
Critically-ill patients often find themselves in the intensive care unit (ICU) seeking specialized support for respiratory distress \cite{doyle1995identification,ware2000acute}. Despite advances in supportive treatments, the in-hospital mortality rate remains 40\% for conditions such as acute lung injury and acute respiratory distress syndrome \cite{rubenfeld2005incidence,sweatt2014evolving}. Managing respiratory distress involves intricate treatment measures, including invasive mechanical ventilation \cite{esteban2000mechanical}, non-invasive mechanical ventilation \cite{esquinas2017noninvasive}, and high-flow nasal cannula \cite{frat2017high}. However, existing recommendations and outcomes, especially regarding intubation and weaning procedures for ICU patients, remain controversial and poorly understood \cite{zuo2020expert,papoutsi2021effect,suo2021machine,wanis2023emulating,kondrup2023towards}.

Health disparities are widespread within marginalized communities, particularly across respiratory diseases, acting as significant contributors to morbidity and mortality in the United States \cite{schraufnagel2013official,moy2017leading,thakur2014lung}. These communities, facing systemic barriers and social inequalities, bear a disproportionate burden of adverse health outcomes due to factors such as economic instability, limited access to education, and housing insecurity \cite{purnell2016achieving}. Recognizing and evaluating social determinants of health (SDOH) is important for addressing the complex factors that influence the quality of and access to healthcare \cite{holmes2023strategies,bundysocial,lua2023effects,marmot2005social,nakagawa2023ai,moukheiber2024looking}. A comprehensive understanding of SDOH can offer insight into potential disparities that might be overlooked within studies focused solely on traditional attributes such as age, race, gender, and health insurance, making it important for the evaluation of algorithmic bias \cite{celi2022sources,nazer2023bias}.

Observational health data, derived from EHRs, presents a valuable resource with the potential to enhance healthcare. Although efforts have been made to establish benchmarks for EHR data \cite{harutyunyan2019multitask,purushotham2018benchmarking,wang2020mimic,gupta2022extensive,rocheteau2021temporal}, these benchmarks primarily focus on conventional clinical prediction tasks, such as mortality and length-of-stay predictions. To the best of our knowledge, the current benchmark datasets lack dynamic aspects of pulmonary function, encompassing complex respiratory treatment strategies,  ventilator settings, and pulmonary mechanics, along with other clinically-relevant variables for guiding decision-making. Furthermore, current ICU benchmark datasets often lack a link to SDOH, which limits the ability to fully understand and address the complexities influencing the recommendations for intubation and weaning in ICU patients. The recently released MIMIC-IV dataset, linked to SDOH features based on patient zip code \cite{yang2023evaluating}, enables detailed fairness assessments of SDOH dimensions. Therefore, we use MIMIC-IV to benchmark clinical respiratory intervention tasks for ICU patients.

In this work, we benchmark two time-dependent clinically-motivated prediction tasks, including successful weaning and prolonged mechanical ventilation. We further evaluate the differences in performance gaps across protected attributes, including age, gender, race, and English proficiency, as well as eight SDOH features. We also release a dataset with hourly intervals to enable benchmarking of respiratory intervention tasks. This dataset is enriched with ventilation data and a wide range of other covariates, including demographics, lab results, measurements, illness severity scores, treatment interventions, and outcome variables. It covers 50,920 patients admitted to the ICU, with records collected over 90 days. This dataset can help address weaning delays and failures, optimize strategies for respiratory support, identify efficiencies in clinical practices, provide decision support to attending physicians regarding intubation decisions in the ICU, and facilitate time-series and reinforcement learning applications.

\section{Methods}
%\vspace{0.3em}

\subsection{DataSet}
%\vspace{0.3em}

\subsubsection{Dataset Overview}
%\vspace{0.4em}

\begin{figure*}[!ht]
  \centering
  % First image
  \includegraphics[width=1\textwidth]{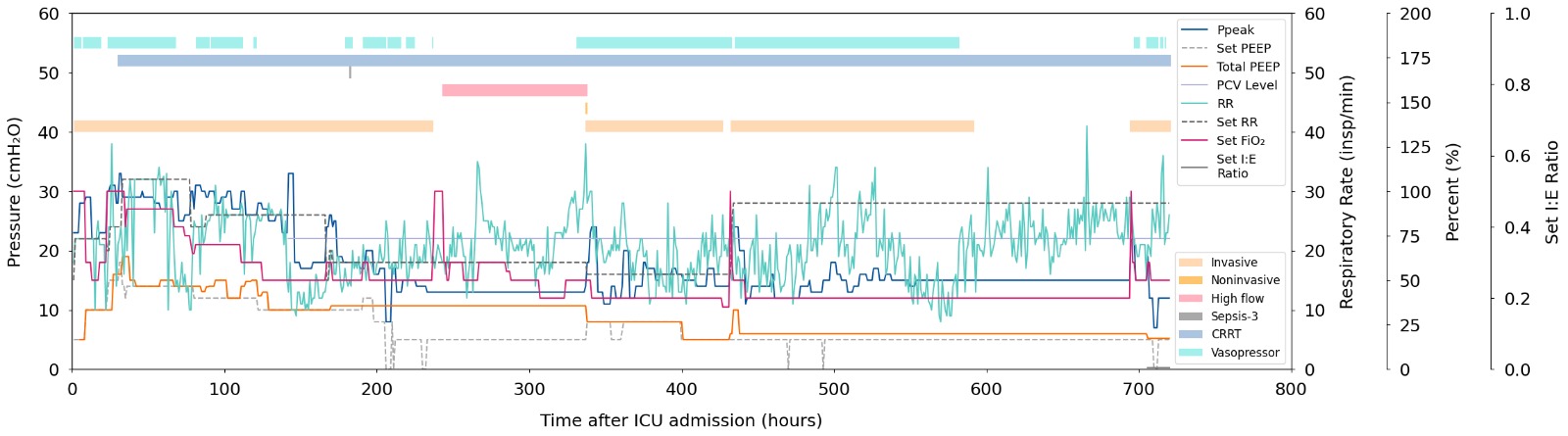}
  
  % Second image
  \includegraphics[width=1\textwidth]{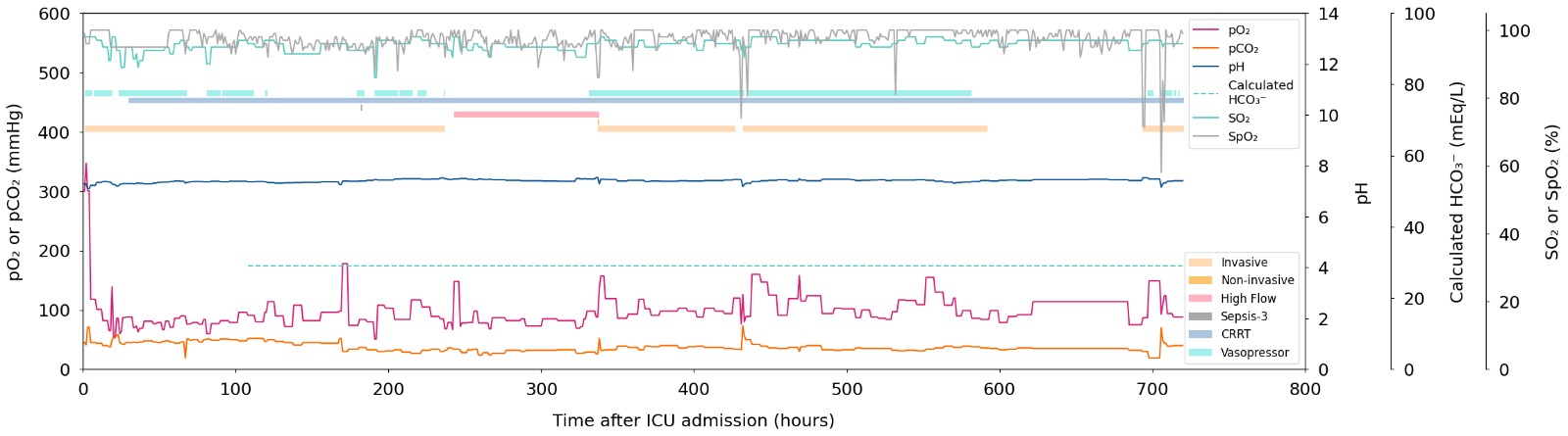}
  
  % Third image
  \includegraphics[width=1\textwidth]{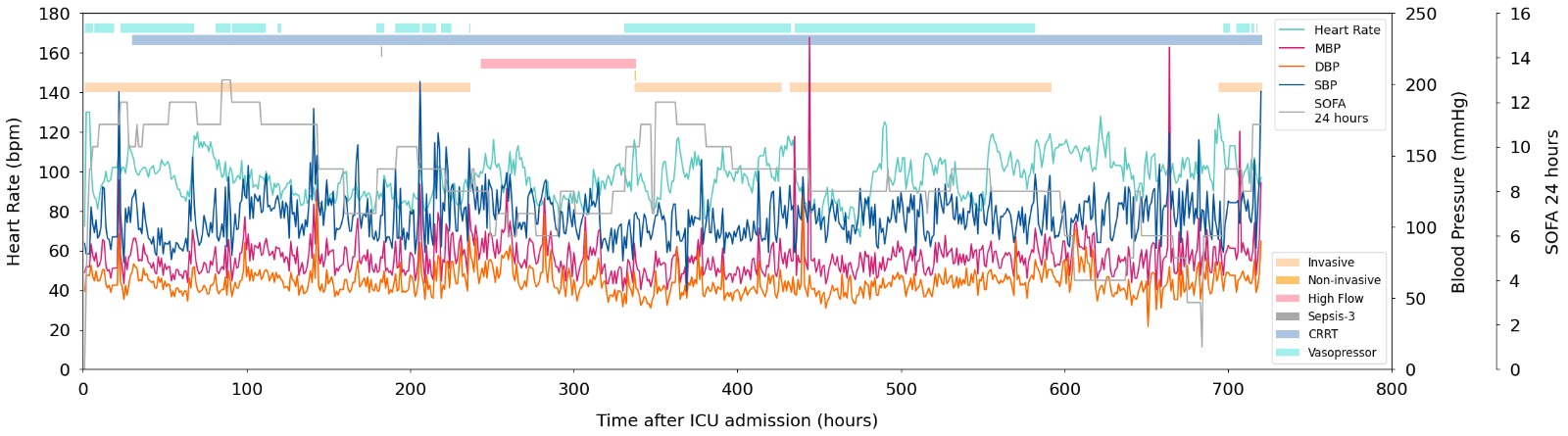}
  
  \caption{Visualization of time-varying covariates for a patient's stay, 30 days after ICU admission. The plots, listed from top to bottom, include ventilation parameters, laboratory results, and vital signs, respectively.
  }
  \label{fig:combined_figures}
\end{figure*}

We introduce a temporal benchmark for clinical respiratory interventions, a 90-day hourly ventilation dataset derived from MIMIC-IV version 2.2. MIMIC-IV is an open-access, de-identified database compiled from electronic health records of patients admitted to the ICU or Emergency Department at the Beth Israel Deaconess Medical Center in Boston between 2008 and 2019. Our temporal data includes confounding variables categorized into static and dynamic variables. Figure\ref{fig:combined_figures} depicts hourly characteristics for a single patient’s ICU stay over 30 days.

\subsubsection{Cohort Selection}
%\vspace{0.4em}

In the MIMIC-IV database, a patient can have multiple stays in the ICU over the years or experience transitions between different ICUs during the same hospital admission. To prevent data leakage and maintain data integrity, we choose the first ICU stay with respiratory support for each patient. This approach ensures that data used for modeling is independent and not influenced by information from subsequent stays. Furthermore, patients without resuscitation or intubation directives and those who were on invasive ventilation 24 hours before admission to the ICU are excluded, resulting in a total of 50,920 patients.

\subsubsection{Data Extraction and Preprocessing}
%\vspace{0.4em}

Most timestamps for the variables that vary over time in the raw MIMIC data are presented in the year, month, day, hour, minute, and second format, offering the potential to derive granular data for comprehensive medical analysis. The sporadic recording of multiple observations allows us to aggregate the data into hourly bins to improve the data density and analytical consistency. Our dataset spans the period of 0 to 2160 hours (equivalent to 90 days) following ICU admission for each subject.

%\vspace{4em}

\subsubsection{Patient-level Static Variables}
%\vspace{0.4em}

The static parameters extracted the patients, as described in Table \ref{tab:static_variables}, include variables such as ICU admission and discharge, sex, type of insurance, language proficiency, marital status, together with clinical details including the patient's first ICU type, predicted body weight in kilograms, height in inches, and the Elixhauser-Van score - a measure of comorbidity.

\begin{table}[!ht]
    %\tiny
    %\small
    %\footnotesize
    \scriptsize
    \centering
    \begin{tabular}{l|l} % Changed {ll} to {|l|l|}
        \toprule
        \textbf{Variable} & \textbf{Description} \\
        \midrule
        intime & ICU admission time \\
        outtime & ICU discharge time \\
        gender & patient gender \\
        anchor year & patient shifted year \\
        anchor age & patient age in anchor year \\
        insurance & patient insurance type \\
        language & English proficiency indicator \\
        marital status & patient marital status \\
        race & patient race \\
        first\_careunit & ICU type during first admission \\
        pbw\_kg & patient predicted body weight (kg) \\
        height\_inch & patient height (inches) \\
        elixhauser\_vanwalraven & Elixhauser-Van Walraven score \\
        \bottomrule
    \end{tabular}
    \caption{\label{tab:static_variables}Patient-level static variables.}
\end{table}

%\vspace{-1em}
\subsubsection{Measurement Observations}
%\vspace{0.4em}

The time-varying measurements in the data encompass ventilation settings, laboratory results, and vital signs, as well as comorbidity scores that evaluate neurological function (including the Glasgow Coma Scale), along with an assessment of patient organ dysfunction (Sequential Organ Failure Assessment score). Ventilation settings and vital signs are extracted from the MIMIC chartevents table, while labs data are obtained from the MIMIC labevents table, each identified by their respective ItemIDs. To handle multiple values within a single hour for a subject, we aggregate the results by computing the median, as the median exhibits reduced sensitivity to noisy data. The labs are sourced from arterial blood gas (ABG) specimens, as arterial blood measurements are deemed to have greater clinical relevance and precision when evaluating parameters such as respiratory function, acid-base balance, and oxygenation status. Two parameters derived from ventilation settings are also presented: set\_pc\_draeger (set pressure for pressure-controlled ventilation from the Draeger ventilator) and set\_pc (set pressure for pressure-controlled ventilation). Set\_pc\_draeger is calculated as the difference between the inspiratory pressure from the Draeger ventilator (pinsp\_draeger) and the set peak inspiratory pressure (ppeak). Based on clinical knowledge, set\_pc  is populated with pcv\_level (pressure controlled ventilation level) if present, pinsp\_hamilton (inspiratory pressure from Hamilton ventilator) if pcv\_level is absent, and set\_pc\_draeger (inspiratory pressure from Draeger ventilator) if both are absent. All variables related to ventilation parameters, vital signs, and labs, and their corresponding descriptions, are described in Table \ref{tab:2}. 
\begin{table}[ht]
    \tiny
    %\scriptsize
    \centering
    \begin{tabular}{l|l}
        \midrule
        \textbf{Variable} & \textbf{Description} \\
        \midrule
        \multicolumn{2}{c}{\textbf{Ventilation parameters}} \\
        \midrule
        ppeak & peak Inspiratory pressure (cmH$_2$O) \\
        set\_peep & set peak inspiratory pressure (cmH$_2$O) \\
        total\_peep & total peak inspiratory pressure (cmH$_2$O) \\
        rr & respiratory rate (insp/min) \\
        set\_rr & set respiratory rate (insp/min) \\
        total\_rr & total respiratory rate (insp/min) \\
        set\_tv & set tidal volume (mL) \\
        total\_tv & total tidal volume (mL) \\
        set\_fio2 & set fraction of inspired oxygen \\
        set\_ie\_ratio & set inspiratory-to-expiratory ratio \\
        set\_pc & set pressure for pressure controlled ventilation (cmH$_2$O) \\
        \_set\_pc\_draeger & set pressure from Draeger Ventilator (cmH$_2$O) \\
        \_pinsp\_draeger & inspiratory pressure from Draeger Ventilator (cmH$_2$O) \\
        \_pinsp\_hamilton & inspiratory pressure from Hamilton ventilator (cmH$_2$O) \\
        \_pcv\_level & pressure controlled ventilation level (cmH$_2$O) \\
        \midrule
        \multicolumn{2}{c}{\textbf{Labs}} \\
        \midrule
        calculated\_bicarbonate & calculated bicarbonate, whole blood (mEq/L) \\
        so2 & oxygen saturation (\%) \\
        pCO2 & partial pressure of carbon dioxide (mmHg) \\
        pO2 & partial pressure of oxygen (mmHg) \\
        pH & pH \\
        \midrule
        \multicolumn{2}{c}{\textbf{Vital Signs}} \\
        \midrule
        heart\_rate & heart rate (bpm) \\
        sbp & systolic arterial blood pressure (mmHg) \\
        dbp & diastolic arterial blood pressure (mmHg) \\
        mbp & mean arterial blood pressure (mmHg) \\
        sbp\_ni & systolic non-invasive blood pressure (mmHg) \\
        dbp\_ni & diastolic non-invasive blood pressure (mmHg) \\
        mbp\_ni & mean non-invasive blood pressure (mmHg) \\
        temperature & temperature (°C) \\
        spO2 & oxygen saturation pulse oximetry (\%) \\
        glucose & blood glucose \\
        \midrule
        \multicolumn{2}{c}{\textbf{Other Variables}} \\
        \midrule
        gcs & Glasgow Coma Scale (GCS) score \\
        gcs\_motor & GCS motor response component \\
        gcs\_verbal & GCS verbal response component \\
        gcs\_eyes & GCS eye-opening response component \\
        gcs\_unable & Endotracheal tube indicator \\
        sofa\_24\_hours & 24-hour Sequential Organ Failure Assessment (SOFA) score \\
        \midrule
    \end{tabular}
    \caption{\label{tab:2}Measurement observations. ``set'' in ventilation settings refers to values set by healthcare professionals on the ventilator to suit the patient’s respiratory needs. ``\_'' refers to intermediate variables.}
\end{table}

%%%%%%%%%%%%%%%%%%%%%%%% Table 2%%%%%%%%%%%%%%%%%%%%%%%%%%%%%%%%%%%%%%

\subsubsection{Treatment Interventions}
%\vspace{0.4em}

% \subsubsection*{Respiratory Support Interventions}
Three respiratory support methods, including invasive ventilation (INV), non-invasive ventilation (NIV), and high-flow nasal cannula (HFNC), are presented as binary indicators per hour. The curation of these respiratory support variables is verified by clinical experts to ensure accuracy and reliability. In MIMIC, the procedureevents table identifies patients on INV or NIV during their ICU stay, while the chartevents table identifies patients on HFNC. INV and NIV in MIMIC have documented start and end times recorded by respiratory therapists, however, HFNC lacks a corresponding time interval; having only the time at which the measurement was observed. Therefore, we pre-process the data to establish a start time and end time for each HFNC event per ICU stay. In addition, multiple HFNC events could occur during a single ICU stay. Therefore, if the time gap between two consecutive HFNC events exceeded 24 hours, we treat them as separate events. For each HFNC event, the minimum and maximum time at which the HFNC is applied is used to obtain the HFNC start time and the end time. HFNC events with identical start and end times are excluded. In cases of overlapping mutually exclusive treatments, where patients are recorded to be on both non-invasive and invasive ventilation simultaneously, we prioritize the most invasive treatment strategy \((\text{INV} > \text{NIV} > \text{HFNC})\). The overlap of mutually exclusive treatments occurs due to the complexities involved in transitioning between ventilation therapies within the ICU, which often includes a series of procedures during the transition period. Furthermore, for short intervals (less than 6 hours) recorded between two different treatments, we attribute the gap to the less invasive treatment. This allows us to handle situations where the precise timing of treatments is unclear. 
% \subsubsection*{2. Additional Interventions}

Additional binary indicators for interventions include vasopressor administration and continuous renal replacement therapy. Vasopressors are extracted from the MIMIC inputevents table and matched to the corresponding hour using their respective start and end times. A patient is classified as being on vasopressors if they received norepinephrine, epinephrine, dopamine, phenylephrine, or vasopressin. Information regarding continuous renal replacement therapy (CRRT) is extracted from the MIMIC chartevents table. Patients are identified as being on CRRT if they have a positive value for blood flow rate or fluid removal during dialysis. A summary of all treatment interventions is presented in Table \ref{tab:3}.

\begin{table}[ht]
    %\tiny
    \scriptsize
    \centering
    \begin{tabular}{l|l}
        \toprule
        \textbf{Variable} & \textbf{Description} \\
        \midrule
        invasive & invasive ventilation indicator \\
        noninvasive & non-invasive ventilation indicator \\
        highflow & high-flow nasal cannula indicator \\
        vasopressor & vasopressor treatment indicator \\
        crrt & continuous renal replacement therapy indicator \\
        \bottomrule
    \end{tabular}
    \caption{\label{tab:3}Treatment interventions.}
\end{table}

%\vspace{-0.5em}

\subsubsection{Outcome Variables}
%\vspace{0.4em}
The majority of the outcome variables are recorded as binary indicators at each hour, with one denoting the occurrence of the event. These include discharge outcome, ICU out-time outcome, death outcome, and sepsis. 
Discharge outcome and ICU out-time outcome indicate if a patient was discharged from the hospital or ICU respectively. The death outcome variable denotes whether a patient died at a specific hour. The date of death in MIMIC is derived from hospital and state records. In cases where both data sources are available, in-hospital mortality is preferentially used over state-linked data. The state-derived date of death includes only the date component, so a default time of midnight is used when converting the date to a timestamp. The data also includes a sepsis outcome variable that identifies whether a patient is septic according to the Sepsis-3 diagnostic criteria. Additionally, it contains the length of stay variable, which indicates the duration of a patient's ICU stay in fractional days. A summary of the outcome variables is presented in Table \ref{tab:4}.

\begin{table}[ht]
%\tiny
    \scriptsize
    \centering
    %\begin{tabular}{p{3cm}|p{10cm}}
    \begin{tabular}{l|l}
        \toprule
        \textbf{Variable} & \textbf{Description} \\
        \midrule
        discharge\_outcome & hospital discharge indicator \\
        icuouttime\_outcome & ICU discharge indicator \\
        death\_outcome & death indicator \\
        sepsis & presence of sepsis using sepsis 3 criteria \\
        los & ICU length of stay (fractional days) \\
        \bottomrule
    \end{tabular}
    \caption{\label{tab:4}Outcome variables.}
\end{table}

% \vspace*{50pt}

%%%%%Data visual%%%%%
%%%%%Data visual%%%%%

%%%%%Data visual%%%%%
%\vspace{4em}

\subsection{Benchmark Tasks}
%\vspace{0.2em}

We consider two clinically motivated prediction tasks for respiratory interventions in ICU settings: prolonged mechanical ventilation and successful weaning.

%\vspace{0.4em}
\noindent
\textit{\textbf{Task definition for prolonged mechanical ventilation:}} Prolonged mechanical ventilation can increase the caregiver burden and affect a patient's quality of life \cite{vali2023prediction,sayed2021predicting}. We aim to predict prolonged mechanical ventilation using the first 24 hours of data in the ICU. Specifically, we define prolonged mechanical ventilation as the initial attempt to ventilate a patient for more than 14 days in the ICU.

%\vspace{0.4em}
\noindent
\textit{\textbf{Task definition for prolonged successful weaning:}} Weaning has been studied in recent clinical trials \cite{pham2023weaning}. In this study, we use the first attempt to separate a patient from a mechanical ventilator. We aim to predict prolonged successful weaning using five days of ICU stay data. Specifically, we define successful weaning as no re-intubation or death within seven days of extubation.

%\vspace{0.1em}

The pre-processed patient cohorts for prolonged mechanical ventilation and successful weaning includes 4,930 and 2,358 cases, respectively. The numerical features for each task are normalized by min-max scaling. For each task, we split the data into 70\% training, 10\% validation, and 20\% testing, while ensuring no patient overlap in the sets to avoid data leakage. In our hybrid sequence-based models, we combine continuous and static features to capture both the hourly dynamics of a patient's condition and the patient's individual characteristics, providing a comprehensive basis for our binary classification tasks on a stay level. For our hybrid fully connected networks which do not involve recurrent connections, we employ static features in conjunction with the median of the time-series features.
%\vspace{0.05em}

\subsubsection{Model Architecture}
%\vspace{0.4em}

\begin{figure}[H]
    \centering
    \includegraphics[width=0.45\textwidth]{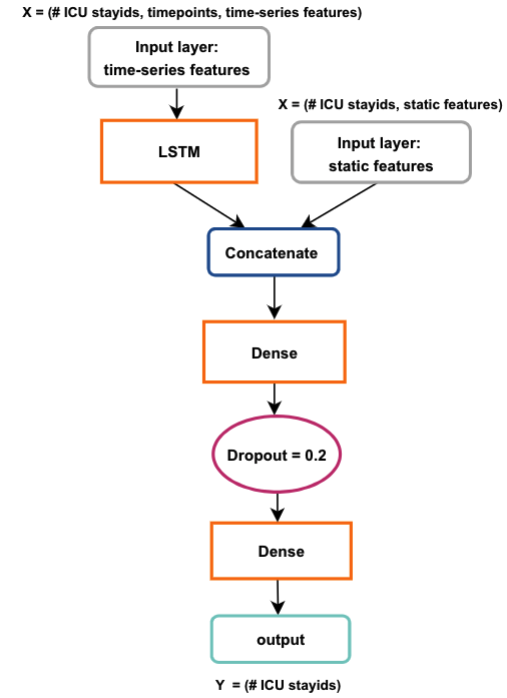}
    \caption{Model architecture of the sequence models. The architecture uses a joint-fusion strategy that concatenates hourly time-dependent features with static features and includes recurrent layers such as an LSTM, BiLSTM, or GRU, or neural networks like TCN.}

   \label{fig:model-arch}
   \end{figure}

In our proposed benchmark, we employ five types of machine learning models to address the aforementioned tasks. We specifically focus on deep learning-based methodologies, including sequence models and a multilayer perceptron (MLP) aiming to assess whether models that operate over time-steps can enhance overall model performance. The sequence models encompass Gated Recurrent Units (GRU), Long Short-Term Memory (LSTM), Bidirectional LSTM (BiLSTM), and Temporal Convolutional Neural Networks (TCN). For the sequence-based models, we apply a joint-fusion strategy to concatenate the hourly time-dependent variables with the static variables. Details of the sequence-based models are depicted in Figure \ref{fig:model-arch}. We fine-tune each model through an exhaustive hyperparameter search. The learning rate is initialized at 0.001 and decays by 5\% for each epoch, with a batch size set to 512. The optimization algorithm used is the Adam Optimizer, and the loss function used is binary cross-entropy. We stop the model training when the validation loss does not improve over three consecutive epochs.

\subsubsection{Model Evaluation}
We assess the models by evaluating their accuracy using the area under the receiver operating characteristic curve (AUROC). To perform binary classification on the predictions, we determine the optimal threshold value by selecting the threshold that maximizes the difference between true positive rate and false positive rate, and then we compare each prediction against this threshold. We report 95\% confidence interval of the evaluation metrics, calculated by performing bootstrapping on the metric scores over 1000 iterations.

\begin{figure*}[t]
    \centering
    \includegraphics[width=1\textwidth,height=8cm]{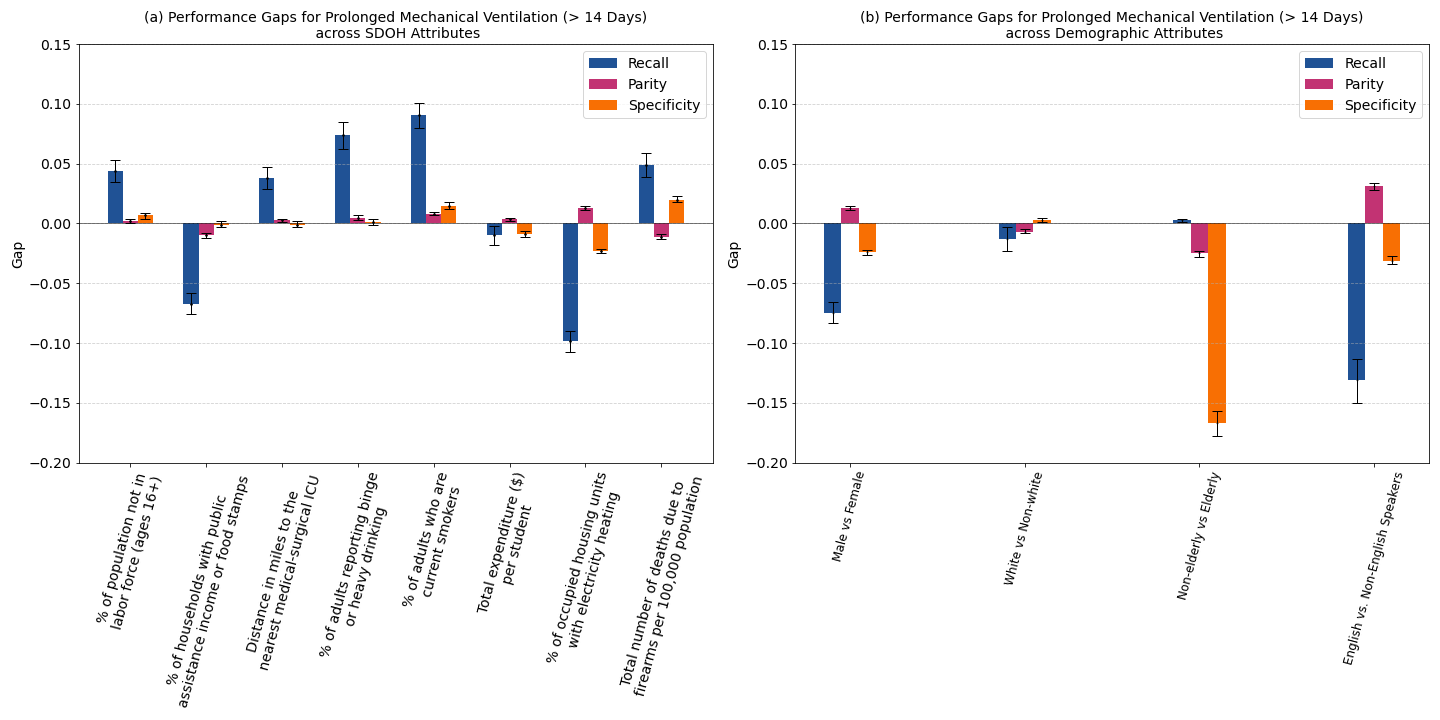}\vfill
    \caption{Performance gap measures for the prolonged mechanical ventilation task under the best model (GRU). A positive bar indicates the model favors one group over the other group. Error bars denote a 95\% confidence interval obtained through 1000 bootstrap samples. \hspace{3em} a) Performance gap evaluation for SDOH attributes. b) Performance gap evaluation for demographic attributes.}
    \label{fig:3}
\end{figure*}

\begin{figure*}[!ht]
    \centering
    \includegraphics[width=1\textwidth,height=8cm]{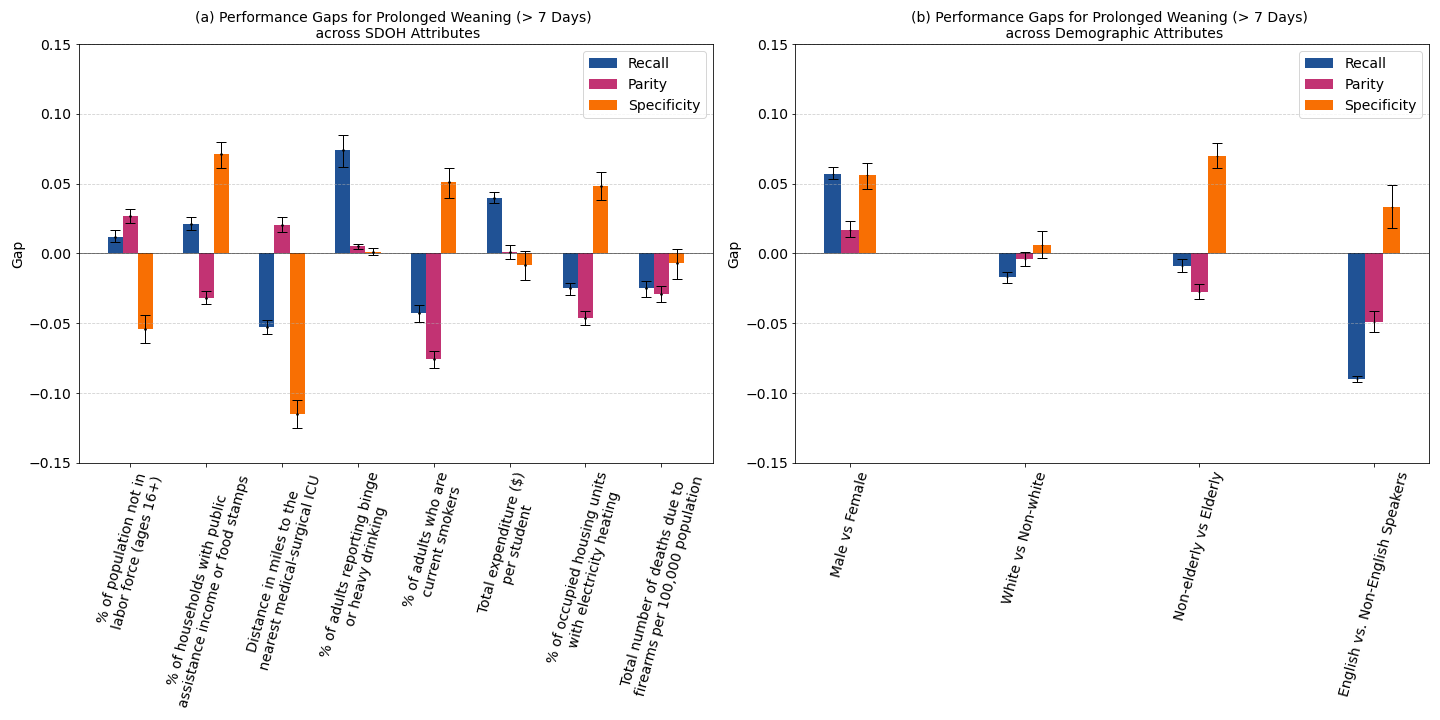}\vfill
    \caption{Performance gap measures for the successful prolonged weaning task under the best model (GRU). A positive bar indicates the model favors one group over the other group. Error bars denote a 95\% confidence interval obtained through 1000 bootstrap samples. \hspace{3em} a) Performance gap evaluation for SDOH attributes. b) Performance gap evaluation for demographic attributes.}
    \label{fig:4}
\end{figure*}

\subsection{Fairness Audits Along SDOH \& Demographic Attributes}
%\vspace{0.2em}

We perform fairness audits by considering protected attributes, such as race, age, and gender, along with eight SDOH attributes. This provides deeper insights into the patient population beyond conventional demographic attributes. We utilize the MIMIC-IV census tract-level SDOH data to conduct fairness audits on our benchmark tasks \cite{yang2023evaluating}. Our analysis includes investigating the differences in fairness across subgroups based on SDOH attributes, such as whether a patient resides in areas with high employment rates, has a high reliance on public assistance or food stamps, lives close to healthcare facilities, engages frequently in heavy drinking or smoking, has high student expenditure, resides in homes with high electricity heating and lives in areas with few deaths from firearms. 

We assess the performance of downstream classifiers based on three definitions of fairness, including, demographic parity (parity gap), equality of opportunity for the positive class (recall gap), and equality of opportunity for the negative class (specificity gap) \cite{chen2019can}. We follow methods used in prior work to expand the demographic parity gap \cite{zhang2020hurtful,hashimoto2018fairness}, and use a similar process to obtain the recall, and specificity gaps. These evaluations are conducted on the best-performing model for both tasks.
%\vspace{-0.1em}

\section{Results \& Discussion}
\subsection{Benchmark Tasks}
%\hspace{-0.3em}
The AUROC for both prolonged mechanical ventilation and weaning are shown in Table \ref{tab:model_metric_mv}. We found that the sequence-based model, the GRU (Hybrid) model, outperforms all other models on both binary prediction tasks.

\begin{table*}[t]
\scriptsize
    \centering
    \begin{tabular}{l|l|l}
        \toprule
        \textbf{Model} & \textbf{Mechanical Ventilation (AUROC ↑)} & \textbf{Successful Weaning (AUROC ↑)} \\
        \midrule
        MLP & 0.641 (0.638 - 0.643) & 0.749 (0.747 - 0.751) \\
        TCN (Hybrid) & 0.747 (0.745 - 0.749) & 0.743 (0.741 - 0.744) \\
        LSTM (Hybrid) & 0.770 (0.768 - 0.772) & 0.764 (0.763 - 0.766) \\
        BiLSTM (Hybrid) & 0.775 (0.773 - 0.777) & 0.752 (0.750 - 0.753) \\
        GRU (Hybrid) & 0.778 (0.776 - 0.780) & 0.776 (0.774 - 0.778) \\
        \bottomrule
    \end{tabular}
    \caption{\label{tab:model_metric_mv}Benchmark results for two clinically-motivated tasks: classifying mechanical ventilation lasting more than 14 days, using 24 hours of data, and successful weaning lasting more than 7 days, using 5 days of data. Scores are reported with 95\% confidence intervals obtained through 1000 bootstrap samples.}
\end{table*}

\noindent
\subsection{Fairness Audits on Benchmark Tasks}

We illustrate the differences in parity, recall, and specificity for demographic and social determinants of health attributes in the mechanical ventilation (Figure \ref{fig:3}) and successful weaning tasks (Figure \ref{fig:4}) using the best performing model (GRU). Recall indicates the proportion of actual positive instances that the model correctly identifies. It is particularly relevant in clinical settings where minimizing false negatives is crucial for timely effective patient diagnosis. To analyze variations in model performance among continuous SDOH attributes, we discretize the attributes into two quantiles. A positive recall gap suggests that the model favors the low prevalence of the specified SDOH attribute over the high prevalence. For categorical variables like gender, race, age, and English proficiency, a positive recall gap indicates that the model favors males, whites, non-elderly individuals, or English speakers over their respective counterparts.

In Figure \ref{fig:3}a,  for the task of predicting prolonged mechanical ventilation the model favors individuals who reside in areas with high employment rate, have a high reliance on public assistance or food stamps, are close to a medical-surgical ICU, rarely engage in heavy drinking or smoking, have high student expenditure, reside in homes with high electricity heating, and live in areas with few deaths from firearms. Additionally, as seen in Figure \ref{fig:3}b the model favors certain demographic groups, including females, non-white individuals, younger individuals, and non-English speakers. On the other hand, as depicted in Figure \ref{fig:4}a, for the task of predicting weaning, the model favors individuals who reside in areas with high employment rate, have a low reliance on public assistance or food stamps, are far from a medical-surgical ICU, rarely engage in heavy drinking, often smoke, have low student expenditure, reside in homes with high electricity heating, and live in areas with more deaths from firearms. 
Additionally, as seen in Figure \ref{fig:4}b, the model favors certain demographic groups, including males, non-white individuals, elderly individuals, and non-English speakers. 

The performance gaps illustrate the disparities in the model's predictive performance and the necessity for fairness auditing prior to model deployment. By assessing SDOH in addition to the previously studied traditional labels we hope to  disentangle biases and uncover other hidden confounders and associations.

\section{Conclusion}
In critical care settings, it is important to carefully assess model biases across demographic and SDOH attributes before deployment. In this study, we benchmark two time-dependent tasks, including successful weaning and prolonged mechanical ventilation. Using different fairness definitions, we evaluate the differences in performance gaps for both tasks across demographic and SDOH attributes. Furthermore, we release an hourly dataset to support the benchmarking of respiratory intervention tasks. Our work aims to enable the development of machine learning models for timely interventions in critical care, emphasizing the consideration of social determinants to promote equitable healthcare access and improve patient outcomes.

\section*{Data Availability}
The temporal dataset for respiratory support in critically ill patients is hosted on PhysioNet \cite{Moukheiber2024}. It is available at this link, \url{https://doi.org/10.13026/0d8j-2w14}. The presented dataset consists of 50,920 distinct adult patients admitted to the ICU of Beth Israel Deaconess Medical Center (Boston, MA, USA) between 2008 and 2019. We extract static, time-varying, and outcome variables from MIMIC-IV in an hourly materialized view and store the content for each patient in a *.csv format named after the patient’s unique identifier (subject ID).

\section*{Code Availability}
We provide the GitHub repository at 
\url{https://github.com/respiratory-support/respiratory-interventions} which includes SQL scripts, offering tools for data management, querying, and analysis. Python scripts are also provided to demonstrate the application of the dataset in various clinical prediction tasks.

\section*{Acknowledgements}
D.M., and L.M. are supported by grant NIH-R01-EB017205 from the National Institute of Health and by the National Institute of Biomedical Imaging and Bioengineering (NIBIB) under NIH-R01-EB030362. D.M. is also supported by NIH National Library of Medicine, and Massachusetts Life Sciences Center. H.C. is supported by the Korea Health Technology Research and Development Project through the Korea Health Industry Development Institute, funded by the Ministry of Health and Welfare, Republic of Korea. The authors would like to thank Dr. Leo Celi, Dr. Kerollos Wanis, and Sicheng Hao for their valuable feedback. 

%\newpage
\bibliographystyle{named}
\bibliography{ijcai24}

\end{document}